\documentclass[twocolumn,aps,prc,floatfix]{revtex4}
\usepackage{graphicx}
\usepackage{dcolumn}
\usepackage{bm}
\usepackage{mhchem}
\begin{document}

\title{Probing the boundary of phase transition of nuclear matter using proton
flows \\ in heavy-ion collisions at 2-8 GeV/nucleon}
\author{Ya-Fei Guo$^{1,2,3}$}
\author{Gao-Chan Yong$^{1,3}$}
\email[]{yonggaochan@impcas.ac.cn}
\affiliation{$^{1}$Institute of Modern Physics, Chinese Academy of Sciences, Lanzhou 730000, China \\
$^{2}$School of Nuclear Science and Technology, Lanzhou University, Lanzhou 730000, China\\
$^{3}$School of Nuclear Science and Technology, University of Chinese Academy of Sciences, Beijing, 100049, China}

\date{\today}

\begin{abstract}

Based on the relativistic transport model ART with the
hadronic equation of state extended to have a phase transition via the use of the
MIT bag model, properties of phase transition of dense nuclear matter formed in relativistic heavy-ion collisions are investigated. Proton sideward and directed flows are calculated with different equation of states in Au + Au collisions at beam energies of 2, 4, 6 and 8 GeV/nucleon. Compared with AGS experimental data in existence, the boundary of first-order phase transition is roughly confined, i.e., in the range of 2.5-4 times saturation density with temperature about 64-94 MeV. Such constraints are useful for ongoing RHIC Beam Energy Scan-II program to study the QCD matter phase diagram.

\end{abstract}

\maketitle

\section{Introduction}

Studying phase transition from nuclear matter to quark-gluon plasma (QGP) is a main purpose of heavy ion collisions at relativistic energies \cite{Gu01,Br03,Br02}. Phase transition is a smooth crossover at finite temperature and small baryon chemical potential according to lattice QCD calculation \cite{Ka01,Be01,Ba01}. At larger chemical potential, it changes to a first-order phase transition, and the endpoint of the first-order phase boundary is named as the critical point \cite{As01,St01,Ca01,Br01}. To understand the phase structure of QCD, it is important to know the position of the critical point and the
phase boundary in the QCD phase diagram. However, the exact value of the transition density from hadronic matter to quark matter is still a matter of long-standing debate in both nuclear physics and astrophysics \cite{Sc01,Sc02,Di01,pt1ellip,pt2scan,pt320,pt4xie}. Heavy ion collisions at relativistic energies are the only practical method to study the QCD phase transition on earth. Both experimental measurements and transport model calculations demonstrate that at alternating-gradient synchrotron (AGS) energies, heavy-ion collisions can form hot and dense matter with densities beyond 3$\rho_0$, and temperatures above 50 MeV \cite{Li01,Li02,tem2006}. There is a large variety of theoretical calculations and experimental measurements in such energy domain dedicating to finding the sign of phase transition \cite{Ba03,Pi01,Li04,Ah01,Ba02,Ch02,Og01,Is01,Ke01,luo17,xu2020,sun17,sh19}.
Unfortunately, up to now, the critical point and the phase-transition boundary still not come to a conclusion.

As a result of phase transition, the Equation of State (EoS) of nuclear matter is soften when changing from hadronic matter to quark matter. Directed nucleon flow caused by directed pressure in semi-central heavy-ion collisions should be affected by the softened EoS. In this study, to search for the phase-transition boundary, we investigate the mean transverse momentum and direct flow of protons in Au + Au collisions at beam energies of 2, 4, 6 and 8 GeV/nucleon in the framework of A Relativistic Transport (ART) model \cite{Li01,Li02} by varying the EoSs with and without phase transition. The first-order phase transition of nuclear matter in Au+Au heavy-ion collisions at 2-8 GeV/nucleon beam energy is deduced to occur at 2.5-4 times saturation density with temperature about 64-94 MeV. The obtained boundary of phase transition here can be verified and further refined by experimental data from the beam energy scan phase II (BES-II) program \cite{pt2scan}.

\section{Model and method descriptions}

The relativistic transport model ART \cite{Li01,Li02} stems from the original Boltzmann-Uehling-Uhlenbeck (BUU) model \cite{Be02,Li03}, but includes more baryons and meson resonances as well as their interactions. More detailedly, the following baryons
$N,~\Delta(1232),~N^{*}(1440),~N^{*}(1535),~\Lambda,~\Sigma$, and
mesons $\pi,~\rho,~\omega,~\eta,~K$ with their explicit isospin
degrees of freedom are included. The model has been successfully used
in studying many features of heavy-ion reactions at AGS energies
up to a beam momentum of about 15 GeV/c \cite{Li02}.
An extended version of ART is now used as a
hadronic afterburner in the AMPT (A Multi-Phase Transport Model)
model for heavy-ion collisions mainly at RHIC and LHC energies
\cite{ampt}.

It is undeniable that the single nucleon potential at high energies (above 1 GeV kinetic energy)
and high densities (above 3 times saturation density) is still unknown to date \cite{mdi2020}.
The density and momentum-dependent single nucleon potential given by \emph{Danielewicz} in Ref.~\cite{mf2000} is applicable only at low energies (below 1 GeV kinetic energy) and low densities (below 2-3 times saturation density). The experimental Hama potential \cite{hama90} at saturation density is of little help in studying the directed flow originating from high baryon densities in heavy-ion collisions at several GeV beam energies. To make minimum assumption, for the single nucleon mean-field potential at hadronic phase (HP), we use the density-dependent form \cite{Ga01}, i.e.,
\begin{eqnarray}
U(\rho)=&&\alpha\frac{\rho}{\rho_0}+
\beta(\frac{\rho}{\rho_0})^\gamma,
\label{potential}
\end{eqnarray}
where $\rho_0$ stands for the saturation density. The parameter values $\alpha$= -232 MeV, $\beta$= 179 MeV and $\gamma$= 1.3 are obtained by fitting the ground state properties of nuclear matter, i.e., the binding energy $E_0$ = -16 MeV, pressure $P_0$= 0 MeV/fm$^3$ and the most probable incompressibility of symmetric nuclear matter $K_0$= 230 MeV at saturation density \cite{Yo01,Zu01,Zu02}.
Energy per nucleon of nuclear matter is thus expressed as
\begin{eqnarray}
E(\rho)&=&\frac{8\pi}{5mh^3\rho}p^5_{f}+\frac{\alpha}{2}\frac{\rho}{\rho_0}
+\frac{\beta}{1+\gamma}(\frac{\rho}{\rho_0})^\gamma.
\label{epern}
\end{eqnarray}
The pressure $P_H$ in hadronic matter is given by
\begin{eqnarray}
P_H&=&\rho^2\frac{\partial E}{\partial\rho}\frac{16\pi}{15mh^3}p^5_{f}+\frac{\alpha}{2}\frac{\rho^2}{\rho_0}
+\frac{\beta\gamma}{1+\gamma}\frac{\rho^{\gamma+1}}{\rho_0^\gamma}.
\label{preh}
\end{eqnarray}
We describe the properties of quark matter with the MIT bag model, and
consider massless $u$ and $d$ quarks \cite{Ch01,Bu01}.
The pressure $P_Q$ in quark matter at zero temperature reads
\begin{eqnarray}
P_Q=\frac{3}{4}\frac{\pi}{(\pi-2\alpha_s)^{1/3}}\rho^{4/3}-B,
\end{eqnarray}
where $B$ = 150 or 90 MeV/fm$^3$ is the bag constant, $\alpha_s=0.1$ is the QCD coupling constant \cite{Li02,Bu01,Ni01}.

\begin{figure}[t]
\centering
\includegraphics[width=0.5\textwidth]{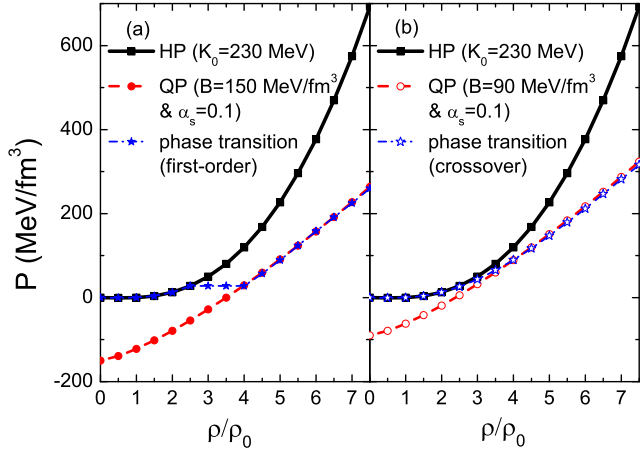}
\caption{(Color online) Pressure as a function of baryon density $\rho/\rho_0$ for hadronic phase (solid line), quark phase (dash line) and mixed phase (dash-dotted line). The fitted EoSs with first-order (a) and crossover (b) phase transitions use respectively B= 150 and 90 MeV/fm$^3$ in the MIT model.}\label{press}
\end{figure}
Figure~\ref{press} shows pressure as a function of baryon density for different phases.
The solid lines denote the pressure $P_H$ for hadronic phase (HP) with $K_0= 230$ MeV, and the dash lines denote the pressure $P_Q$ for quark phase (QP) with $B= 150$ MeV/fm$^3$ (panels (a)) and $B= 90$ MeV/fm$^3$ (panel (b)), respectively. We use the isobaric phase transition at 2.5-4 times saturation density to represent the first-order phase transition from the hadronic phase to the quark phase.
The dash-dotted lines shown in panels (a) and (b) in Figure~\ref{press} stand for the EoSs for the \emph{first-order} and the \emph{crossover} phase transition, respectively. 
In the ART model, these EoSs are obtained from Eq.~(\ref{epern}) or Eq.~(\ref{preh}) by
using the values of $\alpha$, $\beta$ and $\gamma$ given by Eq.~(\ref{potential}) for low densities. For high densities, we use instead
$\alpha = -29.81 - 46.9\frac{K_{0}+44.73}{K_{0}-166.32}$ MeV, 
$\beta = 23.45\frac{K_{0}+255.78}{K_{0}-166.32}$ MeV, 
$\gamma = \frac{K_{0}+44.73}{211.05}$ MeV
to model the EoS of quark matter in terms of the nucleon degrees of freedom. In the
above, the density dependence of the incompressibility parameter $K_{0}(\rho/\rho_{0})$ is
$K_{0}(\rho/\rho_{0}) = 2.324(\rho/\rho_{0})^{3} - 48.23(\rho/\rho_{0})^{2} + 328.54(\rho/\rho_{0}) - 591.46$ for B = 150 MeV/fm$^{3}$ and $K_{0}(\rho/\rho_{0}) = 0.755(\rho/\rho_{0})^{2} - 20.09(\rho/\rho_{0}) + 265.11$ for B = 90 MeV/fm$^{3}$.
Since the chance of the crossover occurs at AGS energies with high baryochemical potential and low temperature is rare \cite{pt2scan}, in the following studies, to probe the signal of phase transition, besides the EoS of crossover, we frequently compare the results obtained by using EoSs with and without first-order phase transition.

\section{Results and discussions}

\begin{figure}[t]
\centering
\includegraphics[width=0.45\textwidth]{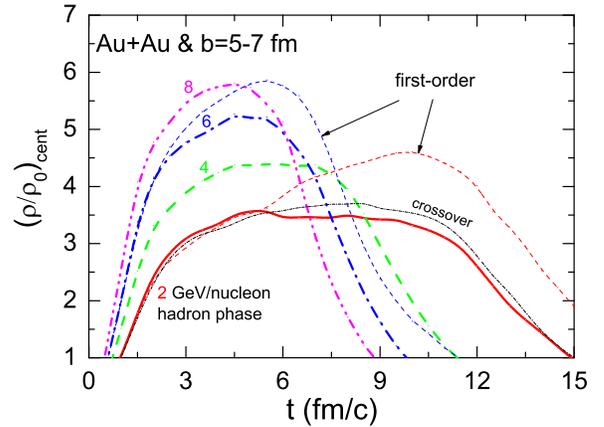}
\caption{(Color online) Maximum compression density reached with hadronic EoS (thick lines) and EoSs including first-order phase transition and crossover (thin lines) in Au + Au collisions at beam energies of 2-8 GeV/nucleon.}\label{den}
\end{figure}
Transverse flow as a probe of nuclear matter EoS is derived from the anisotropic pressure gradient in high energy heavy ion collisions at finite impact parameters \cite{Li05}.
Isobaric phase transition can lead to a soft region in the nuclear matter, which generates a minimum pressure-driven sideward flow. Thus, the change of the strength of sideward flow may be a signal of phase transition. The collective sideward flow in Au + Au collisions at beam energies of 2, 4, 6, and 8 GeV/nucleon has already been measured by the E895 experimental group at Brookhaven \cite{Li04}. In the collisions, as shown in Figure~\ref{den}, the maximum baryon densities reached are approximately 3.5-6 times saturation density. The maximum baryon densities reached become larger when phase transition occurs in nuclear matter. It is interesting to see if enhanced compression would cause stronger nucleon sideward or directed flows.

\begin{figure}[t]
\centering
\includegraphics[width=0.5\textwidth]{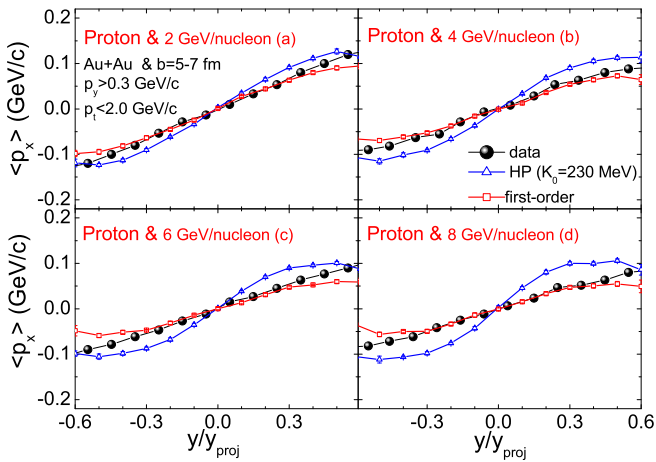}
\caption{(Color online) Proton average transverse momentum $\langle p_x\rangle$ as a function of rapidity in Au + Au collisions using EoSs with and without phase transition at beam energies of 2-8 GeV/nucleon. The experimental data are taken from Ref.~\cite{Li04}.}\label{px}
\end{figure}
Figure~\ref{px} shows the proton average transverse momentum $\langle p_x\rangle$ (sideward flow) given by the ART model with different EoSs, together with the E895 experimental data \cite{Li04}. It is first seen from the proton sideward flow experimental data that, as the incident beam energy increases, the slope of proton flow at mid-rapidity does not increase at all, perhaps indicating occurence of phase transition in dense nuclear matter. Theoretically, in such beam energy range, as increase of incident beam energy, compression density reached in heavy-ion collisions is also enhanced (as shown in Figure~\ref{den}), which would cause a large pressure gradient. Therefore the slope of proton flow should become larger if the interaction time is the same. However, one cannot judge if the main reason is due to the enhanced compression or the shortened interaction time with increasing beam energy. Therefore, one must carry out theoretical simulations by using hadronic transport model. Shown in the four panels of Figure~\ref{px}, one can see that, as increase of incident beam energy, results of proton sideward flow given by the ART model with hadronic EoS gradually deviate from experimental data while results with first-order phase transition are overall well in agreements with experimental data. Figure~\ref{px} indicates the occurence of first-order phase transition in the compression nuclear matter with densities ranging from saturation to 6 time saturation density in heavy-ion collisions at 2-8 GeV/nucleon.

\begin{figure}[t]
\centering
\includegraphics[width=0.5\textwidth]{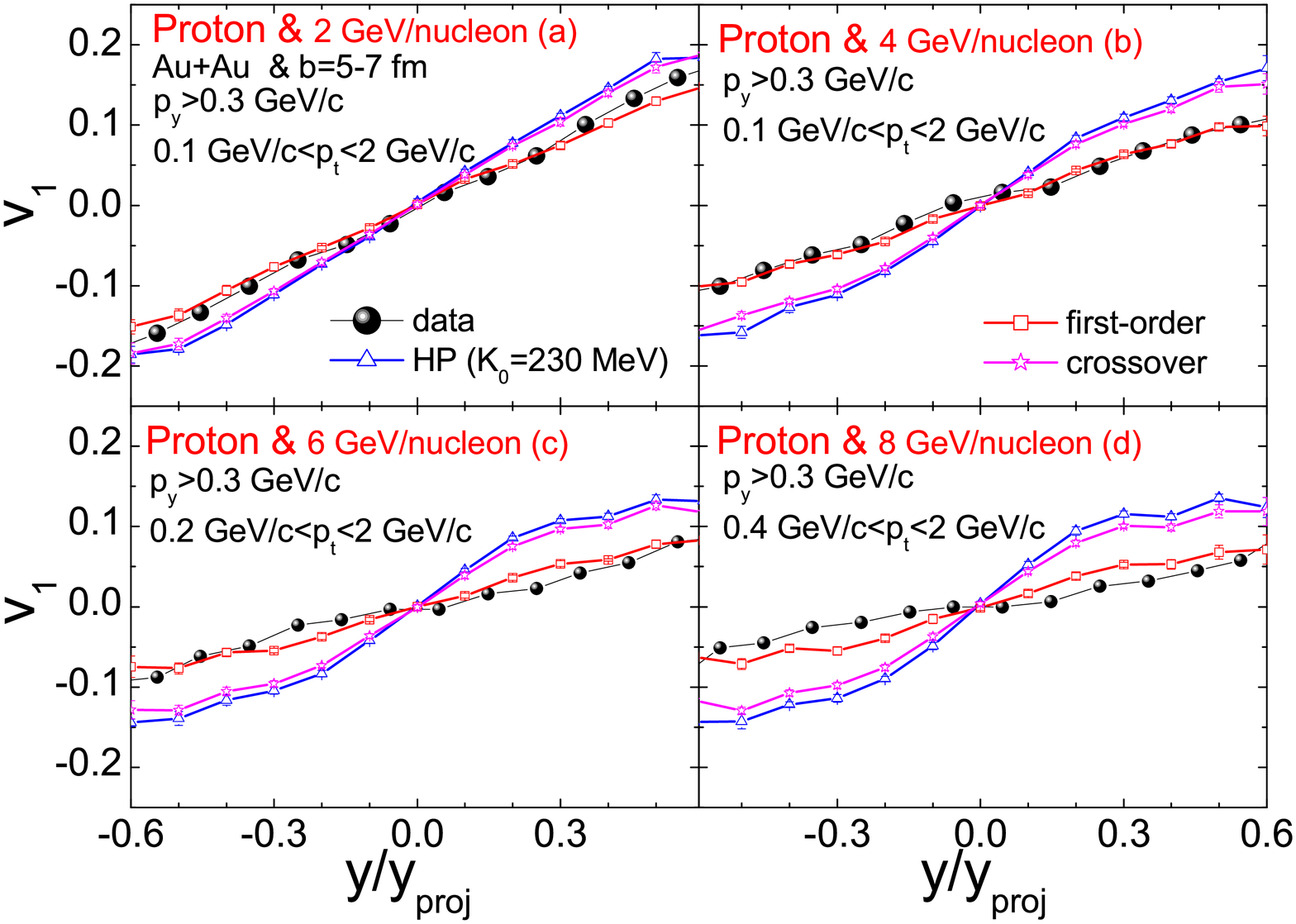}
\caption{(Color online) Proton directed flow $v_{1}$ in Au + Au collisions using three kinds of EoSs at beam energies of 2-8 GeV/nucleon. The experimental data are taken from Ref.~\cite{Li04}.}\label{v_{1}}
\end{figure}
The proton directed flow $v_{1}$ is the first Fourier coefficient of nucleon azimuthal distribution \cite{Ol01}, i.e.,
\begin{equation}
v_{1}=\langle cos\phi\rangle=\langle \frac{p_x}{p_T}\rangle.
\end{equation}
Figure~\ref{v_{1}} shows proton directed flows given by the ART model with different EoSs. One sees that results with hadronic EoS and crossover evidently deviate from experimental data while the results with first-order phase transition fit the data quite well. Compared with experimental data, the deviation of results with hadronic EoS and crossover becomes larger as increase of incident beam energies, indicating larger probability of occurence of first-order phase transition at compression densities above 2.5 times saturation density. The resulting proton directed flows with hadronic EoS and crossover are quite similar, simply because their pressures below 6 times saturation density are close to each other (as shown in Figure~\ref{press}).

\begin{figure}[t]
\centering
\includegraphics[width=0.5\textwidth]{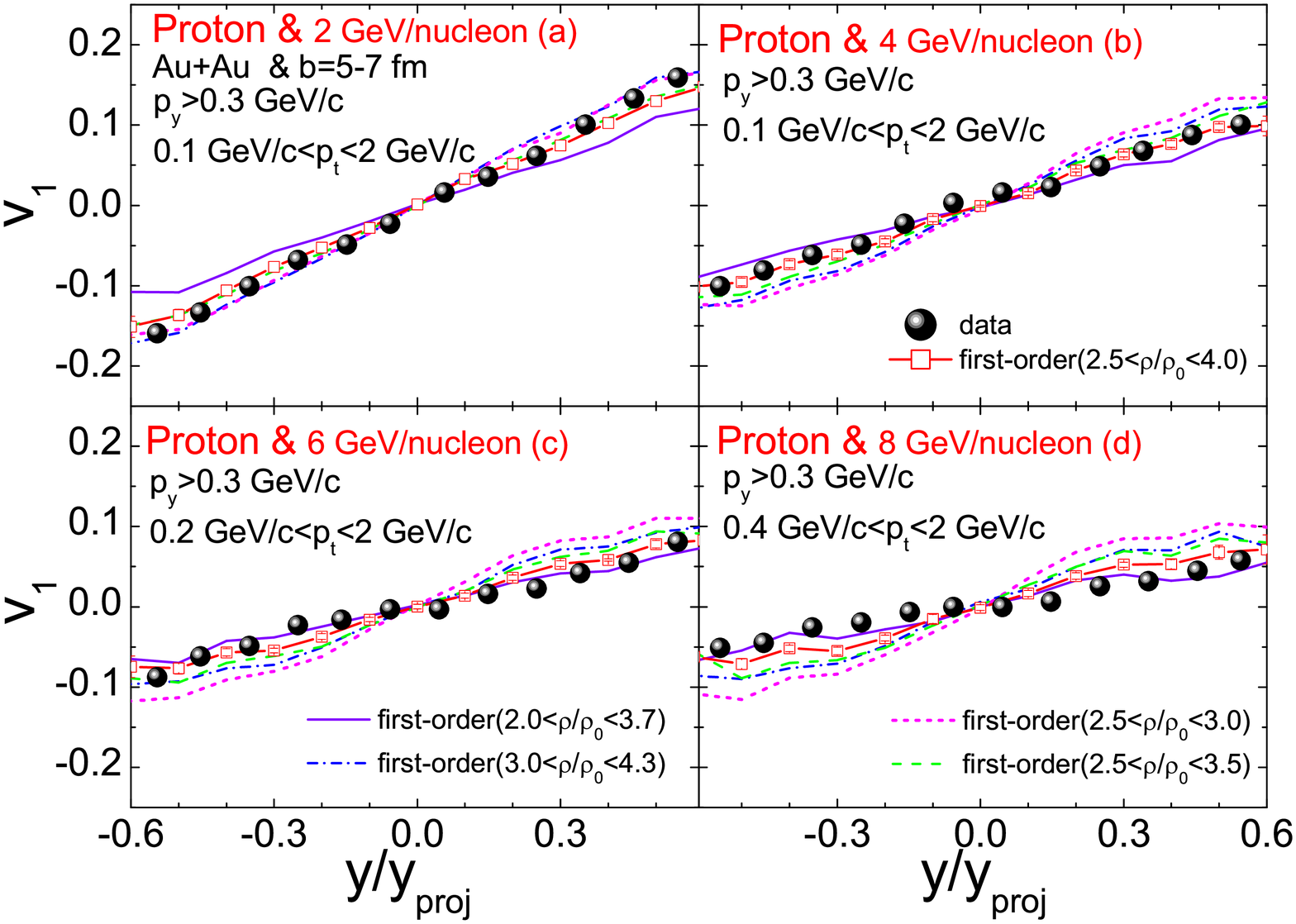}
\caption{(Color online) Same as Fig.~\ref{v_{1}}, but with different phase-transition boundaries. }\label{boundary}
\end{figure}
To probe the boundary of phase transition, one needs to shift the density boundary of isobaric first-order phase transition. From Figure~\ref{boundary}, it is found that, compared to experimental data, our results of proton directed flow become worse at 2-4 GeV/nulcleon beam energies but get improved slightly at 6-8 GeV/nucleon beam energies when using 2$\rho/\rho_{0}$ as the starting density. Whereas with 3$\rho/\rho_{0}$ starting density, all the results do not get improved (technically, to match the MIT bag model calculations, the ending densities are slightly shifted). When altering the ending density of first-order phase transition from 4 to 3$\rho/\rho_{0}$, it is found that the strengths of proton directed flow at 2 GeV/nucleon are almost unchanged, but the slopes of proton directed flow at 4-8 GeV/nucleon are evidently enhanced, becoming far from the experimental data. Shifting the ending density from 4 to 3.5$\rho/\rho_{0}$ causes an apparent discrepancy in proton directed flow at 2 GeV/nucleon, and slight enhancements at 4-8 GeV/nucleon. Setting even larger ending density of first-order phase transition boundary needs larger MIT bag model $B$ parameter than 150 MeV, which seems not to be supported by relevant theories \cite{Li02,Bu01,Ni01}. Therefore, from above discussions, it is deduced that the boundary of first-order phase transition of nuclear matter is around 2.5-4$\rho/\rho_{0}$.

It is highly necessary to ponder whether the in-medium correction of nucleon-nucleon scattering cross section affects the strength or slope of proton directed flow studied here \cite{medcx0}. From Refs.~\cite{medcx1,medcx2,medcx3}, it is see that the in-medium correction of nucleon-nucleon scattering cross is not only density-dependent, but also momentum-dependent, especially at higher momenta. At beam energies of 2-8 GeV/nucleon, the reduced factor of in-medium correction of nucleon-nucleon scattering cross section is expected to approach unity since the nucleon momentum in the nucleon-nucleon center-of-mass system is usually close to or greater than 1 GeV/c \cite{medcx3}. Therefore, the in-medium correction of nucleon-nucleon scattering cross section should have negligible effects on the proton directed flow studied here.

It was argued that the excitation function of elliptic flow in Au+Au collisions at beam energies of 1-11 GeV/nucleon exhibits characteristic signatures which could signal the onset of possible phase transition \cite{pt1ellip}. We therefore investigated the excitation function of elliptic flow in Au+Au collisions at beam energies of 2-8 GeV/nucleon. Although our theoretical calculations overall fit relevant experimental data \cite{Pi01} well, they are unfortunately not sensitive to phase transition.

Finally, it should be pointed out that in Au+Au collisions at 2-8 GeV/nucleon, temperatures reached in formed compression matter are about 64-94 MeV \cite{tem2006}, thus the above discussed boundary of phase transition of nuclear matter should be at such temperature range.

\section{Conclusions}

In summary, within the framework of the relativistic transport model ART with the
hadronic equation of state extended to have a phase transition via the use of the
MIT bag model, the boundary of phase transition of dense nuclear matter is studied using proton sideward and directed flows in Au + Au collisions at beam energies of 2, 4, 6 and 8 GeV/nucleon. The results indicate that a first-order phase transition of dense nuclear matter occurs at densities about 2.5-4 times saturation density with temperature about 64-94 MeV. To more accurately constrain the phase-transition
boundary from experimental data requires better information on the EoSs of nuclear
and quark matters used in the transport approaches. Also, more precise experimental
data from the ongoing BES-II program at RHIC is expected to give better insight into
the QCD matter phase diagram.

The author GCY acknowledges B. A. Li for constructive comments. This work is supported in part by the National Natural Science Foundation of China under Grant Nos. 11775275.


\begin{thebibliography}{99}

\bibitem{Gu01} S. Gupta, X. F. Luo, B. Mohanty, H. G. Ritter, and N. Xu, Science \textbf{322}, 6037 (2011).
\bibitem{Br03} M. A. Braun, J. Dias de Deus, A. S. Hirsch, C. Pajares, R. P. Scharenberg, and B. K. Srivastava, Phys. Rep. \textbf{599}, 1 (2015).
\bibitem{Br02} P. Braun-Munzinger, V. Koch, T. Sch$\ddot{a}$fer, and J. Stachel, Phys. Rep. \textbf{621}, 76 (2016).
\bibitem{Ka01} F. Karsch, E. Laermann, and A. Peikert, Nucl. Phys. B \textbf{605} 579 (2001).
\bibitem{Be01} C. Bernard, T. Burch, C. DeTar, J. Osborn, S. Gottlieb, E. Gregory, D. Toussaint, U. Heller, and R. Sugar (MILC Collaboration), Phys. Rev. D \textbf{71}, 034504 (2005).
\bibitem{Ba01} A. Bazavov \emph{et al.}, Phys. Rev. D \textbf{85}, 054503 (2012).
\bibitem{As01} M. Asakawa and K. Yazaki, Nucl. Phys. A \textbf{504}, 668 (1989).
\bibitem{St01} M. A. Stephanov, Prog. Theor. Phys. Suppl. \textbf{153}, 139 (2004).
\bibitem{Ca01} S. Carignano, D. Nickel, and M. Buballa, Phys. Rev. D \textbf{82}, 054009 (2010).
\bibitem{Br01} N. Bratovic, T. Hatsuda, and W. Weise, Phys. Lett. B \textbf{719}, 131 (2013).
\bibitem{Sc01} K. Schertler, C. Greiner, and M. H. Thorna, Nucl. Phys. A \textbf{616}, 659 (1997).
\bibitem{Sc02} K. Schertler, C. Greiner, J. Schaffner-Bielich, and M. H. Thoma, Nucl. Phys. A \textbf{677}, 463 (2000).
\bibitem{Di01} M. Di Toro \emph{et al.}, Phys. Rev. C \textbf{83}, 014911 (2011).
\bibitem{pt1ellip}P. Danielewicz, Roy A. Lacey, P.-B. Gossiaux, C. Pinkenburg, P. Chung, J. M. Alexander, and R. L. McGrath, Phys. Rev. Lett. \textbf{81}, 2438 (1998).
\bibitem{pt2scan}Tetyana Galatyuk, Nucl. Phys. A \textbf{982}, 163 (2019).
\bibitem{pt320}Agnieszka Sorensen, Volker Koch, arXiv:2011.06635 (2020).
\bibitem{pt4xie}Wen-Jie Xie and Bao-An Li, arXiv:2009.13653 (2020).
\bibitem{Li01} B. A. Li and C. M. Ko, Phys. Rev. C \textbf{52}, 2037 (1995).
\bibitem{Li02} B. A. Li, A. T. Sustich, B. Zhang, and C. M. Ko, Int. J. Mod. Phys. E \textbf{10}, 267 (2001).
\bibitem{tem2006}A. Andronic, P. Braun-Munzinger, J. Stachel, Nucl. Phys. A \textbf{772}, 167 (2006).
\bibitem{Ba03} J. Barrette, R. Bellwied, S. Bennett \emph{et al.}, Phys. Rev. C \textbf{56}, 3254 (1997).
\bibitem{Pi01}C. Pinkenburg, N. N. Ajitanand, J. M. Alexander \emph{et al.}, Phys. Rev. Lett. \textbf{83}, 1295 (1999).
\bibitem{Li04}H. Liu, N. N. Ajitanand, J. Alexander \emph{et al.}, Phys. Rev. Lett. \textbf{84}, 5488 (2000).
\bibitem{Ah01}L. Ahle, Y. Akiba, K. Ashktorab \emph{et al.}, Phys. Lett. B \textbf{476}, 1 (2000).
\bibitem{Ba02}B. B. Back, R. R. Betts, J. Chang, W. C. Chang \emph{et al.}, Phys. Rev. Lett. \textbf{87}, 242301 (2001).
\bibitem{Ch02} P. Chung, N. N. Ajitanand, J. M. Alexander \emph{et al.}, Phys. Rev. C \textbf{66}, 021901(R) (2002).
\bibitem{Og01}C. A. Ogilvie, Nucl. Phys. A \textbf{698}, 3 (2002).
\bibitem{Is01} M. Isse, A. Ohnishi, N. Otuka, P. K. Sahu, and Y. Nara, Phys. Rev. C \textbf{72}, 064908 (2005).
\bibitem{Ke01}G. Kestin and U. Heinz, Eur. Phys. J. C \textbf{61}, 545 (2009).
\bibitem{luo17}X. Luo, N. Xu, Nucl. Sci. Tech. \textbf{28}, 112 (2017).
\bibitem{sun17}K. J. Sun, L. W. Chen, C. M. Ko, and Z. B. Xu, Phys. Lett. B \textbf{774}, 103 (2017).
\bibitem{sh19}Edward Shuryak, Juan M. Torres-Rincon, Phys. Rev. C \textbf{100}, 024903 (2019).
\bibitem{xu2020}A. Bzdak, S. Esumi, V. Koch, J. Liao, M. Stephanov, N. Xu, Phys. Rep. \textbf{853}, 1 (2020).
\bibitem{Be02} G. F. Bertsch and S. Das Gupta, Phys. Rep. \textbf{160}, 189 (1988)
\bibitem{Li03} B. A. Li, W. Bauer, and G. F. Bertsch, Phys. Rev. C \textbf{44}, 2095 (1991).
\bibitem{ampt}Z. W. Lin, C. M. Ko, B. A. Li, B. Zhang, S. Pal, Phys. Rev. C \textbf{72} (2005) 064901.
\bibitem{mdi2020}Yasushi Nara, Tomoyuki Maruyama, and Horst Stoecker, Phys. Rev. C \textbf{102} (2020) 024913.
\bibitem{mf2000} P. Danielewicz, Nucl. Phys. A \textbf{673}, 375 (2000).
\bibitem{hama90}S. Hama, B. C. Clark, E. D. Cooper, H. S. Sherif and R. L. Mercer, Phys. Rev. C \textbf{41}, 2737 (1990).
\bibitem{Ga01} C. Gale, G. M. Welke, M. Prakash, S. J. Lee, and S. Das Gupta, Phys. Rev. C \textbf{41}, 1545 (1990).
\bibitem{Yo01} D. H. Youngblood, H. L. Clark, and Y. W. Lui, Phys. Rev. Lett. \textbf{82}, 691 (1999).
\bibitem{Zu01} S. Shlomo, V.M. Kolomietz, G. Col\`{o}, Eur. Phys. J. A \textbf{30}, 23 (2006).
\bibitem{Zu02} J. Piekarewicz, J. Phys. G \textbf{37}, 064038 (2010).
\bibitem{Ch01} A. Chodos, R. L. Jaffe, K. Johnson, C. B. Thorn, and V. F. Weisskopf, Phys. Rev. D \textbf{9}, 3471 (1974).
\bibitem{Bu01} G. F. Burgio, M. Baldo, P. K. Sahu, and H. J. Schulze, Phys. Rev. C \textbf{66}, 025802 (2002).
\bibitem{Ni01} O. E. Nicotra, M. Baldo, G. F. Burgio, and H.-J. Schulze, Phys. Rev. C \textbf{74}, 123001 (2006).
\bibitem{Li05} B. A. Li and C. M. Ko, Phys. Rev. C \textbf{58}, 1382(R) (1998).
\bibitem{Ol01}J. Y. Ollitrault, Nucl. Phys. A \textbf{638}, 195c (1998).
\bibitem{medcx0}Bao-An Li, Lie-Wen Chen, Phys. Rev. C \textbf{72}, 064611 (2005).
\bibitem{medcx1}Qingfeng Li, Zhuxia Li, Sven Soff, Marcus Bleicher and Horst St\"{o}cker, J. Phys. G: Nucl. Part. Phys. \textbf{32}, 407 (2006).
\bibitem{medcx2}Ying Yuan, Qingfeng Li, Zhuxia Li, and Fu-Hu Liu, Phys. Rev. C \textbf{81}, 034913 (2010).
\bibitem{medcx3}Wen-Mei Guo, Gao-Chan Yong, Yongjia Wang, Qingfeng Li, Hongfei Zhang, Wei Zuo, Phys. Lett. B \textbf{726}, 211 (2013).

\end{thebibliography}
\end{document}